# Lattice compression increases the activation barrier for phase segregation in mixed-halide perovskites

Loreta A. Muscarella[1], Eline M. Hutter[1,2], Francesca Wittmann[1], Young Won Woo[3], Young-Kwang Jung[3], Lucie McGovern[1], Jan Versluis[1], Aron Walsh[3,4], Huib J. Bakker[1], Bruno Ehrler[1,*]

[1] Center for Nanophotonics, AMOLF, Science Park 104, 1098 XG Amsterdam, the Netherlands

[2] Department of Chemistry, Utrecht University, Princetonlaan 8, 3584 CB, Utrecht, the Netherlands

[3] Department of Materials Science and Engineering, Yonsei University, Seoul 03722, Korea

[4] Department of Materials, Imperial College London, London SW7 2AZ, United Kingdom

**Corresponding Author ***

b.ehrler@amolf.nl

**Abstract**

The bandgap tunability of mixed-halide perovskites makes them promising candidates for light emitting diodes and tandem solar cells. However, illuminating mixed-halide perovskites results in the formation of segregated phases enriched in a single-halide. This segregation occurs through ion migration, which is also observed in single-halide compositions, and whose control is thus essential to enhance the lifetime and stability. Using pressure-dependent transient absorption spectroscopy, we find that the formation rates of both iodide- and bromide-rich phases in MAPb(Br$_x$I$_{1-x}$)$_3$ reduce by two orders of magnitude on increasing the pressure to 0.3 GPa. We explain this reduction from a compression-induced increase of the activation energy for halide migration, which is supported by first-principle calculations. A similar mechanism occurs when the unit cell volume is reduced by incorporating a smaller cation. These findings reveal that stability with respect to halide segregation can be achieved either physically through compressive stress or chemically through compositional engineering.

**TOC**

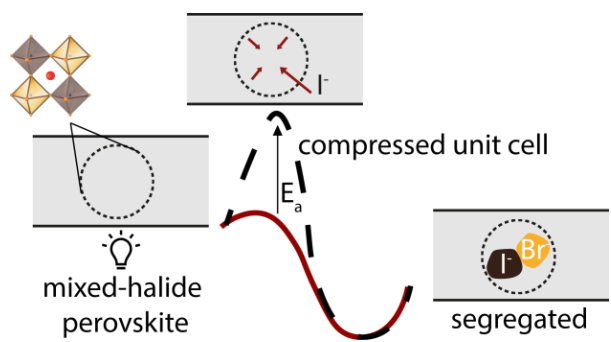

**Introduction**

Halide perovskite semiconductors have recently gathered significant attention due to their excellent optoelectronic properties combined with low-cost and simple fabrication method. In addition, the bandgap of these perovskites can easily be tuned by mixing halides in different ratio[1], making them promising candidates for light emitting diodes (LEDs)[2,3] and tandem solar cells[4]. However, methylammonium(MA)-based mixed-halide perovskites, MAPb(Br$_x$I$_{1-x}$)$_3$, lack long-term bandgap stability under standard solar cell operating conditions[5]. Compositions with $x > 0.20$ have been shown to suffer from halide segregation[6] during continuous light exposure[7] or applied bias[8]. As a result, low- and high-bandgap phases are formed, corresponding to iodide-rich and bromide-rich domains, respectively[6]. Halide migration affects the homogeneity of the bandgap required for many applications such as lighting and displays, and photogenerated charges transfer into the low-bandgap iodide-rich domains, where they recombine.

Phase segregation of mixed-halide perovskites involves the migration of halide ions. In fact, even single-halide perovskite systems (*e.g.* MAPbI$_3$) suffer from ion migration[9,10], and this process has been identified as one of the main drivers for degradation[11]. The activation energies reported for halide migration in single-halide systems[10,12] and in mixed-halide systems[6,12] are comparable, suggesting a similar transport mechanism in both cases. Therefore, both single- and mixed-halide perovskites would benefit from increasing the activation energies for ion migration, as this would slow down the degradation rate and hence increase the lifetime of the corresponding device. Phase segregation has been shown to depend on various factors such as light intensity[12,13,14], duty cycle[15], and film quality and thickness[14,16]. Reducing the halide vacancy concentrations[17] and passivating electron traps[18] whose electric field has been proposed to initiate halide migration by accumulating holes, have been suggested as ways to reduce the rate of phase segregation. The same result has also been achieved by partial replacement of the organic cation with the smaller $Cs^+$ in the mixed-perovskite[19,20,21,22].

Here, we investigate the dependence of the kinetics of phase segregation on structural parameters in mixed-halide perovskite $MAPb(Br_xI_{1-x})_3$ by using pressure-dependent transient absorption spectroscopy. The measurements are performed at hydrostatic pressures ranging from 0 (ambient pressure) to 0.3 GPa using an additional light beam to induce phase segregation. Previously we established that pressure changes the thermodynamic landscape for phase segregation by finding the final mixing ratio of the segregated phases to be closer to the one of the mixed-phase upon increasing pressure[23]. Here, we find that phase segregation is also substantially slower at high pressure for all $MAPb(Br_xI_{1-x})_3$ mixing ratios ($x = 0.25$, 0.5 and 0.7). We explain this slowing down from an increase of the activation energy ($E_a$, in light) for halide migration upon compression of the unit cell volume. Theoretical calculations corroborate this interpretation, revealing an increase in the energy barrier ($\Delta H$) required for a halide species to diffuse into a vacancy under pressure. In addition, we find that compression of the unit cell volume at ambient pressure by partial replacement of the $MA^+$ cation with the smaller $Cs^+$, $e.g.$ $MA_{0.7}Cs_{0.3}Pb(Br_{0.5}I_{0.5})_3$ and $CsPb(Br_{0.5}I_{0.5})_3$, also leads to slower segregation, in a similar manner to the reduced segregation rate of $MAPb(Br_{0.5}I_{0.5})_3$ under high-pressure conditions. Altogether, these findings suggest that compressive strain in the unit cell achieved through compositional engineering or physical pressure can be effectively used to delay halide migration by increasing the activation barrier of the migration process.

**Result and Discussions**

MAPb(Br$_x$I$_{1-x}$)$_3$ thin films with $x = 0$, 0.25, 0.5, 0.7 and 1 were prepared by spin coating the precursor solutions onto quartz substrates as reported elsewhere[23]. Transient absorption spectroscopy (TAS) allows us to probe the excited state of MAPb(Br$_x$I$_{1-x}$)$_3$ by recording absorption spectra of the probed area at each delay time (from 0.05 ps to 800 ps) following a pulsed pump excitation (400 nm, with excitation density of ~10$^{18}$ absorbed photons/cm$^3$ corrected for the fraction of absorbed photons by each composition). **Figure 1a** shows a 2D plot of the transient absorption signal of MAPb(Br$_{0.5}$I$_{0.5}$)$_3$ as a function of the delay time between pump and probe and the probe pulse energy. In absence of an additional light-source, MAPb(Br$_{0.5}$I$_{0.5}$)$_3$ shows a ground state bleach at 2.05 eV in agreement with the absorption onset obtained from steady-state absorption measurements previously reported in Hutter et al[23]. We then focus a continuous wave (CW) light source ($\lambda$ = 405 nm, $I$ = 2.37 × 10$^3$ mW/cm$^2$) on the pump spot to induce phase segregation. In **Figure 1b** we show the 2D transient absorption signal of MAPb(Br$_{0.5}$I$_{0.5}$)$_3$ after 20 minutes of light-soaking . Now, we observe two bleaching signals, one at lower energy (1.85 eV) and one at higher energy (2.15 eV), that can be assigned to iodide- and bromide-rich domains, respectively. At later times, the bleaching signal from the bromide-rich phase decays, indicating the transfer of photogenerated charges into the iodide-rich phase.

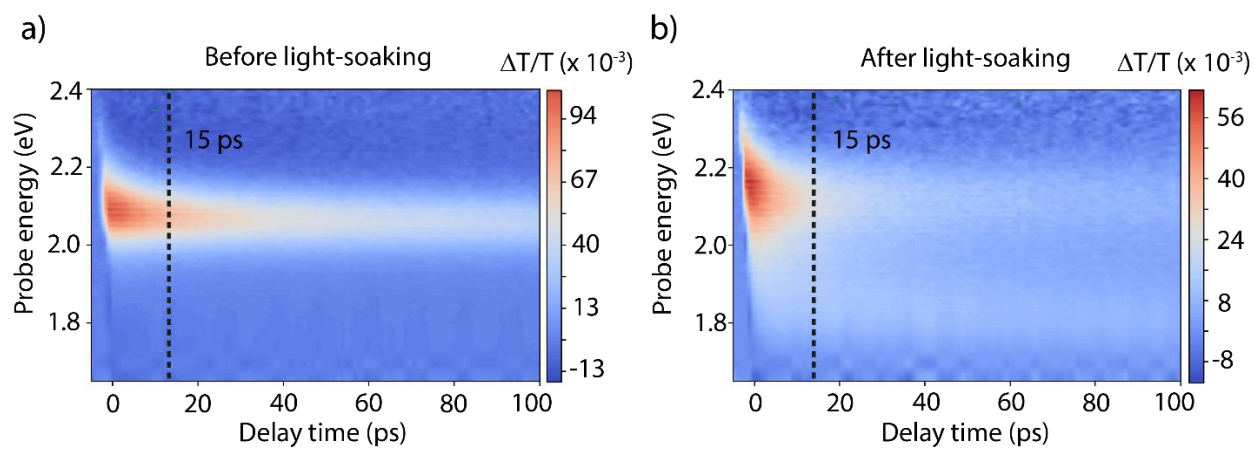

**Figure 1. a)** $\Delta T/T$ as a function of the probe energy and delay time between the pump and probe pulse for MAPb(Br$_{0.5}$I$_{0.5}$)$_3$ before light-soaking where only the bleach from the mixed-halide composition ($x = 0.5$) is observed and **b)** after 20 minutes of light-soaking with a CW-laser where a bleach from two phases peaking at 1.85 eV and 2.15 eV are observed, indicative of halide segregation. The measurements are performed at ambient pressure. The dashed line indicates the delay time chosen to study the kinetics of phase segregation.

To investigate the kinetics of the phase segregation in MAPb(Br$_{0.5}$I$_{0.5}$)$_3$ during light-soaking, we fixed the delay stage position at 15 ps, when most of the hot charge carriers induced by the high-energy pump have cooled[24] to the band edges of the high- and low-energy phases. An important advantage of TAS with respect to time-resolved photoluminescence (TRPL) in this context is the ability to track the excited population from both the iodide- and bromide-rich phase[13], whereas TRPL mainly probes the emissive low-energy phase. We recorded transient absorption spectra every second for 20 minutes. When only the pump excitation is present, we do not observe any phase segregation (see **Figure S1**). When the mixed-halide perovskite film was subjected to CW light irradiation at ambient pressure, we observe the appearance of a second peak at lower energy after a few seconds, which subsequently grows in amplitude. At the same time, the original bleach shifts to higher energy.

Subsequently, we investigated the segregation as a function of compressive strain, which reduces the unit cell size, by performing pressure-dependent transient absorption measurement of the sample inside a hydrostatic pressure cell filled with inert liquid (FC-72, see Experimental Methods). The change in hydrostatic pressure is achieved by increasing the amount of inert liquid through a manual pump, with a resulting pressure ranging from 0 (ambient pressure) to 0.3 GPa. The change in unit cell volume calculated by using the bulk moduli[23] from ambient pressure to 0.3 GPa is about 2.5% for $x = 0.5$ (3% and 2.2% for $x = 0.25$ and 0.7, respectively). The dynamic ingrowth of the iodide- and bromide-rich phases during light-soaking is shown in **Figure 2a**. As pressure increases, we observe a substantial slowing down of the phase segregation. This observation holds true for all MAPb(Br$_x$I$_{1-x}$)$_3$ compositions studied, *e.g.* $x = 0.25$ (which barely segregates from 0.1 GPa onward) and $x = 0.7$, as reported in **Figure S2**. To extract the segregation rate as a function of pressure, Gaussian profiles are fitted to the ground state bleaches of the low- and high-energy peak position in energy, and we monitor the peak energies of these Gaussians as a function of light-soaking time. All kinetic traces are characterized by a continuous change in the ground state bleach energy, due to a change in composition, followed by saturation at a certain energy when segregation is complete. As reported in our previous work[23], the segregated peaks are closer in energy at high pressure, which is attributed to a change in the terminal *x*-value as a function of the unit cell volume. In addition, we observe that the time at which this final composition is reached changes substantially with pressure. The traces were fitted with mono-exponential curves, $Ae^{k_{seg}t} + c$ (eV, offset), to determine the segregation rate $k_{seg}$. Our rates at ambient pressure are comparable to previous work calculated under comparable light-soaking conditions[13,25]. Under pressure, the iodide-rich phase formation rate decreases by almost two orders of magnitude. For $x = 0.5$, $k_{seg}$ decreases from $(0.08 \pm 0.01) s^{-1}$ at ambient pressure to $(0.003 \pm 0.001) s^{-1}$ at 0.3 GPa (**Figure 2b**). A similar trend is observed for $x = 0.7$ and $x = 0.25$, where the segregation rate decreases from $k_{seg} = (0.1 \pm 0.03) s^{-1}$ at ambient pressure to $k_{seg} = (0.003 \pm 0.001) s^{-1}$ at 0.3 GPa, and from $k_{seg} = (0.028 \pm 0.002) s^{-1}$ to $k_{seg} = (0.004 \pm 0.001) s^{-1}$, respectively. For $x = 0.25$, there was no detectable change in the bromide-rich peak within our resolution. For the other compositions, the formation rate of the bromide-rich phase is slower and less affected by pressure compared to the iodide-rich one (**Figure 2b**). This observation will be discussed in more detail later.

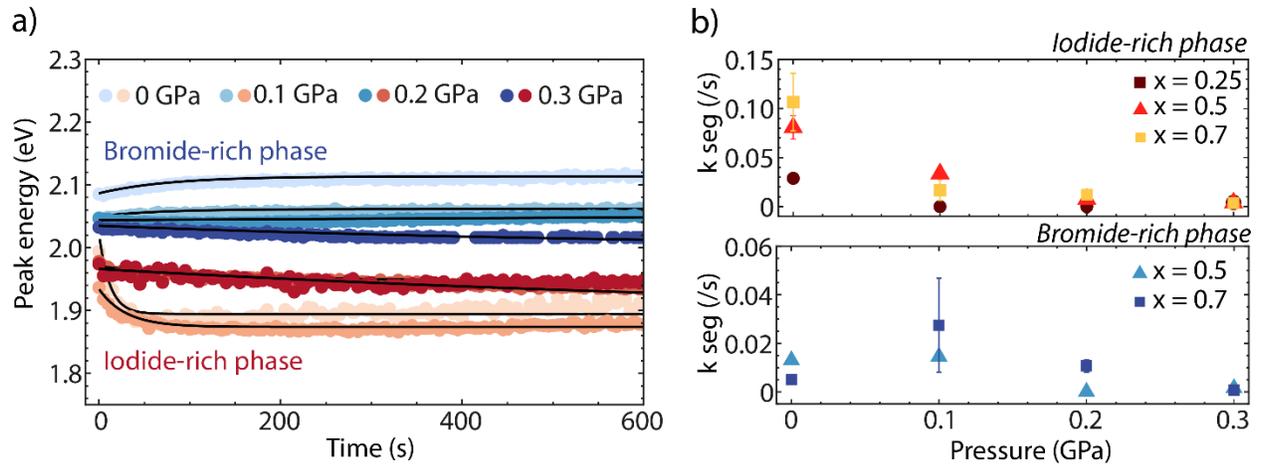

**Figure 2. a)** Dynamic evolution of the low-energy iodide-rich phase (red) and high-energy bromide-rich phase (blue) measured by pressure-dependent TAS during light-soaking with the CW-laser from ambient pressure to 0.3 GPa. **b)** Segregation rates for the iodide- and bromide-rich phase as function of pressure for all the MAPb(Br$_x$I$_{1-x}$)$_3$ compositions ($x = 0.25$, $x = 0.5$ and $x = 0.7$).

We explain the diminution of the segregation rate upon increasing pressure from an increase of the activation energy associated with the migration of halide ions, consequently delaying the accumulation of ions in the low-energy phase. Density functional theory calculations on the model systems $CsPbI_3$ and $CsPbBr_3$ in **Figure 3a** show the energy barrier associated with vacancy-assisted halide diffusion. The potential energy surface for ion diffusion is calculated, and the saddle point is identified, as a function of applied pressure. The transition state increases in energy as the cell volume decreases. As a result, the energy barrier increases with pressure for both phases, with a change of 0.064 eV (I) and 0.056 eV (Br) when the pressure is increased from 0 to 0.8 GPa.

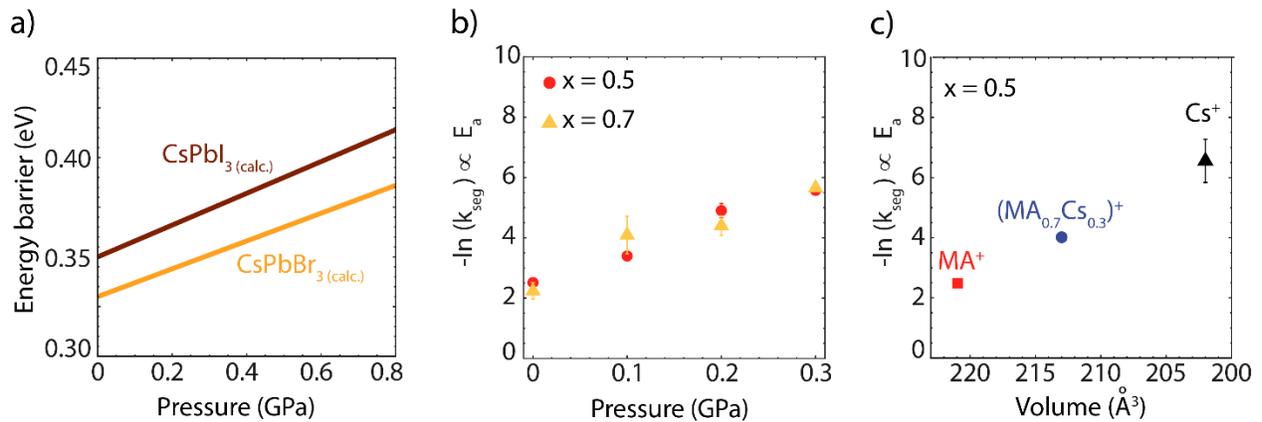

**Figure 3. a**) Calculated (DFT/PBEsol) activation energy for iodide and bromide diffusion in $CsPbI_3$ and $CsPbBr_3$ as a function of pressure. **b**) $-ln(k_{seg})$ (which is proportional to the activation energy for phase segregation), plotted against the external physical pressure applied for $x = 0.5$ and $x = 0.7$, and **c**) against the unit cell volume for $x = 0.5$ where the physical pressure is replaced by chemical pressure *via* partial replacement of $MA^+$ with the smaller $Cs^+$. Here, $-ln(k_{seg})$ follows the same trend as the high-pressure conditions. We observe that $-ln(k_{seg})$ follows the same trend as when the pressure is varied.

To derive the relation between the segregation rate and the activation energy for ion migration we use the definition of the diffusion coefficient[26] (see **Supplementary Note 1** for the full derivation)

$$k_{migration} = \frac{q^2 N v_0 d^2}{6 \epsilon_0 \epsilon_{pero} k_B T} exp\left(\frac{\Delta^\ddagger S°}{k_B}\right) exp\left(-\frac{E_a}{k_B T}\right) \quad (Eq.\ 1)$$

Where $\Delta^\ddagger S°$ is the changes in the entropy of a single ion migration step, $k_B$ is the Boltzmann constant, $v_0$ is the attempt frequency for an ion to hop, $d$ is the ionic hopping distance, $E_a$ is the activation energy, $q$ is the elementary charge, $\epsilon_{pero}$ is the permittivity of the perovskite, $\epsilon_0$ is the permittivity in vacuum and $N$ is the doping density. These terms can be approximated (see **Supplementary Note 1** for discussion) such that the natural logarithm of the migration rate, which in our experiments correspond to $k_{seg}$, is proportional to the activation energy for the migration process as follows:

$$-\ln(k_{migration}) \propto E_a \quad (Eq.\ 2)$$

For simplicity, we define the experimental $-\ln(k_{seg})$ as the effective activation energy $E_a^*$. Typical activation energies for halide migration in these perovskites are in the order of 100-200 meV (see Transient Ion Drift (TID) measurements below)[12,25,27]. The experimental $E_a^*$ plotted as a function of physical pressure applied increases by a factor three on increasing pressure to 0.3 GPa for all the compositions (**Figure 3b).** Interestingly, we show in **Figure 3c** that a similar unit cell compression can be achieved by partial or complete replacement of MA$^+$ with the smaller cation Cs$^+$, corresponding to chemical compression instead of physical pressure (see **Figure S3** for peak energy versus time in the chemically compressed compositions). This observation suggests that slower ion migration is a phenomenon that is more generally linked to a reduction of the unit cell volume, independent of how this reduction is achieved (see **Figure S4** for $E_a^*$ as a function of unit cell volume induced by chemical and physical pressure). In earlier work we found that this generalization is also true for the thermodynamic stability of these mixed halide perovskites[23], pointing towards the importance of the unit cell volume for both kinetic and thermodynamic properties of ion migration. We note here that partial replacement of MA$^+$ with the smaller Cs$^+$ may result in different defect densities and crystallinity, which will also affect the absolute segregation rate. By applying external physical pressure, we essentially vary only one parameter while moving from MA$^+$ to Cs$^+$ may also change other parameters. However, the similarity in the trends observed in **Figure 3b** and **Figure 3c** indicates that the unit cell volume plays a key factor in the phase segregation rate.

Ion migration plays a large role in the degradation of perovskite thin films, and the effect of strain shimmers through many observations related to perovskite stability. For example, residual tensile strain in perovskite films resulting from the thermal expansion coefficient (TEC) mismatch of perovskites with commonly used substrates[28] can result in a tensile stress of 50 MPa within the film[29] after fabrication. Under tensile strain, the inorganic [PbX$_6$]$^{4-}$ framework is distorted resulting in longer and weaker Pb−X bonds and less strongly tilted octahedra[30]. This strain results in a decreased formation energy for defects and a lower activation energy for ion migration[31]. An additional source of tensile strain which lowers the activation energy for ion migration[31,32,33] is the presence of light which has been shown to lead to thermal expansion[34] MA-based mixed halide perovskite unit cells. We also find a lower activation energy under light exposure than under dark conditions, which we measured using TID on an $x = 0.20$ sample, the composition with the highest mixing ratio which does not segregate under light-soaking. We found two negatively charged species migrating with comparable activation energy in dark, namely (0.15 ± 0.03) eV and (0.14 ± 0.01) eV. In single-halide composition only one negative species was observed[10] so it is likely that the two species found in the mixed-halide compositions represent both iodide and bromide. In light, the activation energy of the former species is reduced to (0.09 ± 0.01) eV, whereas the latter remains constant at (0.13 ± 0.02) eV. The decrease in activation energy in light consistent with the observation that light-induced tensile strain leads to increased ion migration. We note that some of the change in activation energy could be due to heating but it cannot explain the full magnitude[35]. Since light changes the activation energy of only one of the two halide migration processes, we propose that the process with the lower activation energy in light may be mainly responsible for phase segregation (see **Supplementary Note 2** and **Figure S7** for TID traces and details on the fitting method).

Our observation that the rate of phase segregation strongly depends on the physical pressure demonstrates the large role that strain plays for ion migration. We hypothesize that pressure-induced compressive stress counteracts the effect of thermal expansion induced by light. Therefore, by strengthening the Pb-X bonds, pressure mitigates the reduction of the activation energy for halide migration. As a result, we expect the total tensile strain at high pressure to be lower and the activation energy for the halide migration to be higher compared to ambient conditions. A scheme of the proposed mechanism at the microscopic and macroscopic level is shown in **Figure 4a** and **Figure 4b**, respectively. A similar effect can be obtained at ambient pressure by inducing compressive strain *via* either regulating tensile strain in the perovskite film through charge-transport layers as recently reported by Xue *et al.*[36] or through compositional engineering, by partial replacement of MA$^+$ with smaller cations. In fact, mixed-cation mixed-halide perovskite compositions have less residual tensile stress and are reported to be less sensitive to light-induced thermal expansion[35] and phase segregation[20,21,22]. As a consequence, Cs$^+$ incorporation further retards phase segregation in MA-based mixed halide perovskites[37].

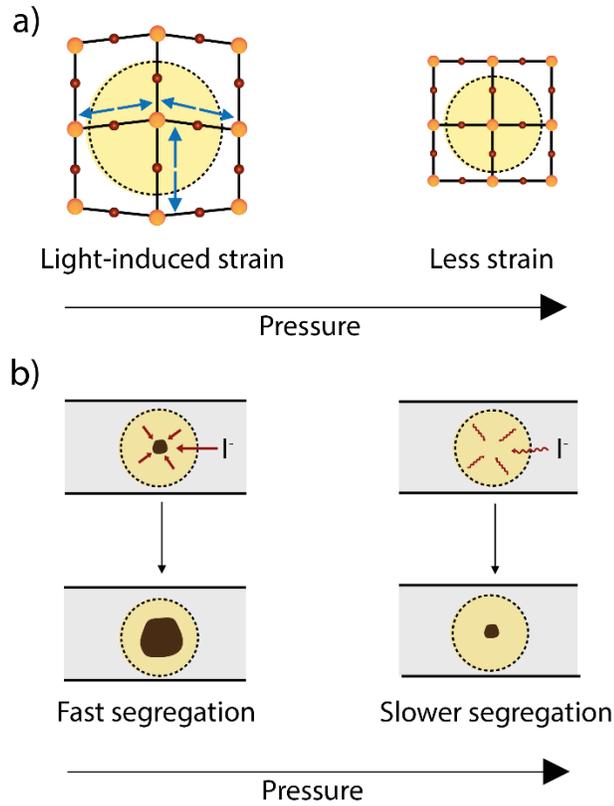

**Figure 4.** Schemes of the proposed mechanism to kinetically stabilize phase segregation by external physical pressure. **a)** Light induces thermal expansion strain and decreases the energy barrier for halide migration by weakening the Pb-X bonds. At high pressure, the compressive strain counteracts the light-induced strain and increases the activation energy for halide migration. **b)** At ambient pressure, mainly the iodide ions move to the illuminated areas (dashed circle) to increase the volume of quickly formed iodide-rich islands, increasing their volume. At high pressure, halide ions move slower given the higher activation energy, leading to slower phase segregation.

Our measurements can also give insight into the mechanism of phase segregation. The formation rate of the low-energy iodide-rich phase appears to be faster than the high-energy bromide-rich phase for all the compositions and pressures explored. The total mixing ratio before light-soaking at ambient conditions calculated from the area and the energy position of the bleach is $0.531 \pm 0.001$ for $x = 0.5$. Now that we know the $x$-values of the segregated phases calculated from the bleach energy positions, and their relative peak areas, we can calculate whether the total mixing ratio has changed during the segregation. After 20 minutes of light-soaking, the mixing ratio is $0.46 \pm 0.01$, suggesting iodide-enrichment of the illuminated area. Although most of the halide ions move within the illuminated area, the decrease in the total mixing ratio suggests that some iodide ions have moved from outside the illuminated region. At high pressure, a similar trend is observed ($x = 0.47 \pm 0.01$ after light-soaking). This similarity in total mixing ratio suggests that despite the changes in segregation rate and final composition, the total number of iodides moving into the probed area does not change with pressure. Furthermore, while both the segregation rate and the final composition of the iodide-rich phase are heavily affected by physical pressure[23], the bromide-rich phase does not show a clear trend with pressure. As an additional observation, the rate of evolution of the area of the bleach of the low-energy phase (roughly proportional to the volume fraction) is slower as compared to the rate obtained by the peak energy position in time, whereas the high-energy phase shows the same rate for the peak energy and peak area decay (see **Figure S5**). The rate of the evolution of the iodide peak area can be ascribed to two competing processes regulating the change in the area of the peak, from one hand it increases due to diffusion of iodide to the illuminated areas of the sample and on the other hand, it decreases because of reduction in the volume due to loss of bromide species. We postulate that all these observations are consistent with the formation of local, small iodide-rich islands (low-energy phase) with a well-defined composition from a small portion of the film which does not substantially affect the high-energy phase. Then, once those islands are formed, the bulk of the halides segregate via the migration of iodide ions into the initially-formed islands, growing the volume of these domains[38]. Increasing pressure, the rates extracted from the peak energy and the peak area come closer until coinciding. We attribute this trend to the reduction of the light-induced tensile strain, which therefore increases the activation energy for halide migration and the formation energy of these islands under illumination with consequent reduction of the phase segregation rate. It is important to note that the different rates extracted from the areas can also be the result of different dynamics before 15 ps for the low- and high-energy phase. Therefore, the interpretation is not trivial. In addition, we observe that during halide segregation, the total area of the two bleaches over time (see **Figure S6**). The total area is constant at high pressure, again likely due to the halide immobilization. Finally, given that the halide migration likely occurs *via* a similar vacancy-mediated mechanism both in mixed- and single-halide compositions, we propose that strain is equally important for halide migration in single-halide systems, and that a similar chemical approach as in mixed-halide compositions could be used to reduce the rate of halide migration in single-halide compositions, *e.g.* MAPbI$_3$.

**Conclusions**

To conclude, we used pressure-dependent transient absorption to investigate the dynamics of phase segregation by tracking the iodide- and bromide-rich phase formation over time. We have shown that phase segregation in several mixed-halide MAPb(Br$_x$I$_{1-x}$)$_3$ ($x$ = 0.25, 0.5 and 0.7) becomes substantially slower under pressure. We attribute the reduced segregation rate to a pressure-induced increase in the effective activation energy for the halide migration process. First-principle calculations support this explanation. We show that chemical compression of the unit cell achieved *via* partial or complete replacement of the $MA^+$ cation by the smaller $Cs^+$ cation results in a comparable slowing down of the phase segregation. Hence, this observation suggests that chemically tuning the unit cell at ambient conditions (instead of applying physical pressure) may be used to kinetically suppress ion migration/phase segregation. Mechanistically, our findings suggest the local formation of small, iodide-rich domains that precede the growth of these islands leading to the final phase segregation. While kinetic stabilization alone may not be a feasible route for long-term stability[39], these findings will help in understanding the key factors affecting the rate of halide migration and to develop an effective strategy to suppress halide migration for the entire lifetime of the perovskite-based devices.

**Supporting Information**

$\Delta T/T$ scan as a function of time and probe energy for $x = 0.5$ without light-soaking at ambient pressure; Evolution of the low-energy iodide- and bromide-rich phase in real time for $x = 0.25$ and $0.7$ and $x = 0.5$ in case of chemical compression; Effective activation energies as a function of unit cell volume induced by chemical and physical pressure; Evolution of the low-energy iodide- and bromide-rich phase area in real time for $x = 0.25$, $0.5$ and $0.7$ and resulting segregation rates; Decay of the total area as a function of pressure for $x = 0.5$ and $0.7$; Derivation of relation between the segregation rate and the activation energy for ion migration; TID measurements and Arrhenius plot for $x = 0.20$ in dark and light;

**Acknowledgments**

The work of L.A.M., E.M.H., F.W., L.McG., J.V., H.J.B. and B.E. is part of the Dutch Research Council (NWO) and was performed at the research institute AMOLF. The work of L.A.M. and L.McG was supported by NWO Vidi grant 016.Vidi.179.005. The work of Y.W.W., Y.K.J. and A.W. was supported by the Creative Materials Discovery Program through the National Research Foundation of Korea (NRF) funded by Ministry of Science and ICT (2018M3D1A1058536). Y.W.W., Y.K.J. and A.W. are grateful to the UK Materials and Molecular Modelling Hub for computational resources, which is partially funded by EPSRC (EP/P020194/1).

**Author contributions**

L.A.M performed the pressure-dependent TA experiments and data analysis together with E.M.H., under the supervision of B.E. F.W. assisted in the sample preparation and J.V. assisted in the TA experiments under the supervision of H.J.B. L.McG. performed the TID experiment and the data analysis. Thermodynamic calculations were performed by A.W, Y.W. and Y.J. The manuscript was written by L.A.M with input from all other authors.